# Twist-Controlled Modulation of Quantum Emitters in a Van der Waals Bilayer

Angus Gale[1*], Seungjun Lee[2*], Seungmin Park[3], Evan Williams[1], Helen Zhi Jie Zeng[1], James Liddle-Wesolowski[1], Young Duck Kim[3], Milos Toth[1,5], Tony Low[2,4], Igor Aharonovich[1,5,†]

1. School of Mathematical and Physical Sciences, University of Technology Sydney, Ultimo, New South Wales 2007, Australia
2. Department of Electrical and Computer Engineering, University of Minnesota, Minneapolis, Minnesota 55455, USA
3. Department of Physics, Kyung Hee University, Seoul, 02447 Republic of Korea
4. Department of Physics, University of Minnesota, Minneapolis, Minnesota 55455, USA
5. ARC Centre of Excellence for Transformative Meta-Optical Systems, Faculty of Science, University of Technology Sydney, Ultimo, New South Wales 2007, Australia

\* These authors contributed equally to this work
† Correspondence to igor.aharonovich@uts.edu.au

**Abstract**

*Stacking and twisting two dimensional materials has garnered enormous attention across the condensed matter and the nanophotonic communities. The surge of interest stems from the emergence of novel photophysical phenomena that arise due to the interlayer coupling of the individual layers. Here, we demonstrate that the twist degree of freedom can modulate a single quantum emitter at room temperature. We employ a van der Waals homobilayer of hexagonal boron nitride (hBN) and model the emission properties of quantum emitters as a function of the twist angle. Density functional theory results show that the embedded emitters are strongly influenced by the twist angle and the stacking of the top hBN layer. We consequently engineer these systems experimentally, and demonstrate in-situ tuning of embedded quantum emitters by mechanically twisting the top hBN layer, achieving tunability of over 30 nm (~ 100 meV). Our work demonstrates that mechanical twisting can be harnessed to modulate the embedded quantum emitters in a vdW material, marking a crucial step towards a programmable on-chip quantum circuitry.*

Twisted bilayers of two dimensional (2D) materials have facilitated a diverse range of physical phenomena[1-6] At particular "magic" angles, bilayers of graphene have been shown to exhibit the formation of flat bands, superconductivity[7, 8] as well as ferromagnetism.[9] These studies have further inspired engineering of moiré graphene superlattices to control and manipulate plasmon resonances at the nanoscale.[10] More recently, moiré systems and have been adapted in other atomically thin materials, including transition metal dichalcogenide (TMDC), where interlayer excitons, correlated systems and Mott insulators have been studied.[11-14]

The unique aspect of the twist degree of freedom is that it could easily be applied for thicker van der Waals (vdW) crystals. As an example, controlling the relative twist interfaces of molybdenum trioxide ($MoO_3$) have also unlocked adaptive and reconfigurable dispersion of phonon polaritons.[15] Hexagonal boron nitride (hBN) is another vdW crystal that has recently been attracting an increased attention due to its potential in a plethora of applications in quantum optics, nanophotonics and optoelectronics.[16-18]

Uniquely to hBN, the stacking order of bulk crystals can result in a number of different properties including a modified band gap[19, 20], emergence of self trapped excitons[21] and enhanced optical nonlinearity.[22] Precise control over twisted hBN interfaces[23] offers an entirely new library of physical properties, such as periodic moiré potential modulation[24, 25] with charge polarisation[26] and even ferroelectric domains.[3] While certain optical applications of twisted hBN have been explored, predominantly at the ultraviolet spectral range near the hBN bandgap, there were no attempts to modulate quantum emitters using a twist degree of freedom.[21, 27, 28] The challenge stems from the fact that dynamic twist of top layers after stacking is extremely challenging due to vdW adhesion between the layers at most angles.[23] Indeed, most previous works report on separate devices that are then compared, rather than a single pair of hBN flakes that can be deterministically tuned. Given the appeal of hBN as a vdW host of high quality single photon emitters (SPEs), it is a prime candidate to explore the possibility of twist based tuning of quantum emitters.

In this work, twisted hBN crystals (referred to as mono-bilayers) are proposed as a model system to enable twist-controlled modulation of quantum emitters. We begin by conducting first principles density functional theory (DFT) calculations of a quantum emitter, and outline the expected energy shifts based on the localised stacking order and twist angle, $\theta_t$. While the origin of the majority of quantum emitters in hBN is still under debate, the carbon trimer ($C_2C_N$ or $C_2C_B$) defects were strong contenders to explain the visible emission in hBN.[29, 30] These emitters are easy to engineer, operate at room temperature and are extremely bright, which is why we focus on them in the current work. We then experimentally realised the mechanical twist and measured the response of the quantum emitters. Remarkably, we could observe spectral shifts of ~ 30 nm (~ 100 meV) from the same SPE, marking it the first twist controlled modulation of any quantum system at room temperature.

Figure 1a shows a schematic of single-photon emission from a carbon trimer color center embedded in hBN. As a wide bandgap (~ 6 eV) vdW crystal, hBN can host a variety of colour centres stemming from localised lattice defects, resulting in discrete electronic energy levels within the band-gap[31, 32]. Since the Bohr radius of a defect-induced exciton in hBN is typically around 1 nm[29, 33], the light emission from the color centers is intuitively sensitive to its local atomic environment. In a pristine hBN lattice, this local environment is uniform with typical bulk material stacked in an AA' configuration, where alternating B (green) and N (grey) atoms are arranged vertically, as shown in Figure 1b. A rotation of one sheet by 60 ° retains the vertical alignment but results in AA stacking order with the same atoms atop one another. Alternatively, when the hBN sheets are shifted in-plane by one bond length, the resulting stacking order is referred to as AB'/BA', or AB/BA if the sheets were already rotated by 60 °. At any angle in between, the twisted bilayers of hBN result in a moiré superlattice, with periodic modulation of stacking order.[34, 35] Figure 1b summarizes the representative stacking configurations in twisted bilayer hBN.

Stacking two hBN layers with a relative twist angle, $\theta_t$ creates a moiré superlattice, which spatially modulates the local atomic configurations and may provide a mechanism to tune the SPEs embedded in the hBN bilayer. Starting from the AA' stacking configuration, a twist introduces BA' and AB' configurations, which are characterized by vertical B-B and N-N stackings, respectively. Similarly, a twist from AA configuration (or near a 60 ° rotation about the hexagon's center) forms local environments with vertical N-B and B-N stackings, which are referred to as BA and AB configurations, respectively. The local BA and AB stacking configurations form out-of-plane electric dipole moments due to their broken inversion symmetry[3, 36], as highlighted with red circles in Figure 1b, while all other local stacking configurations lack a net electrical dipole moment.[35]

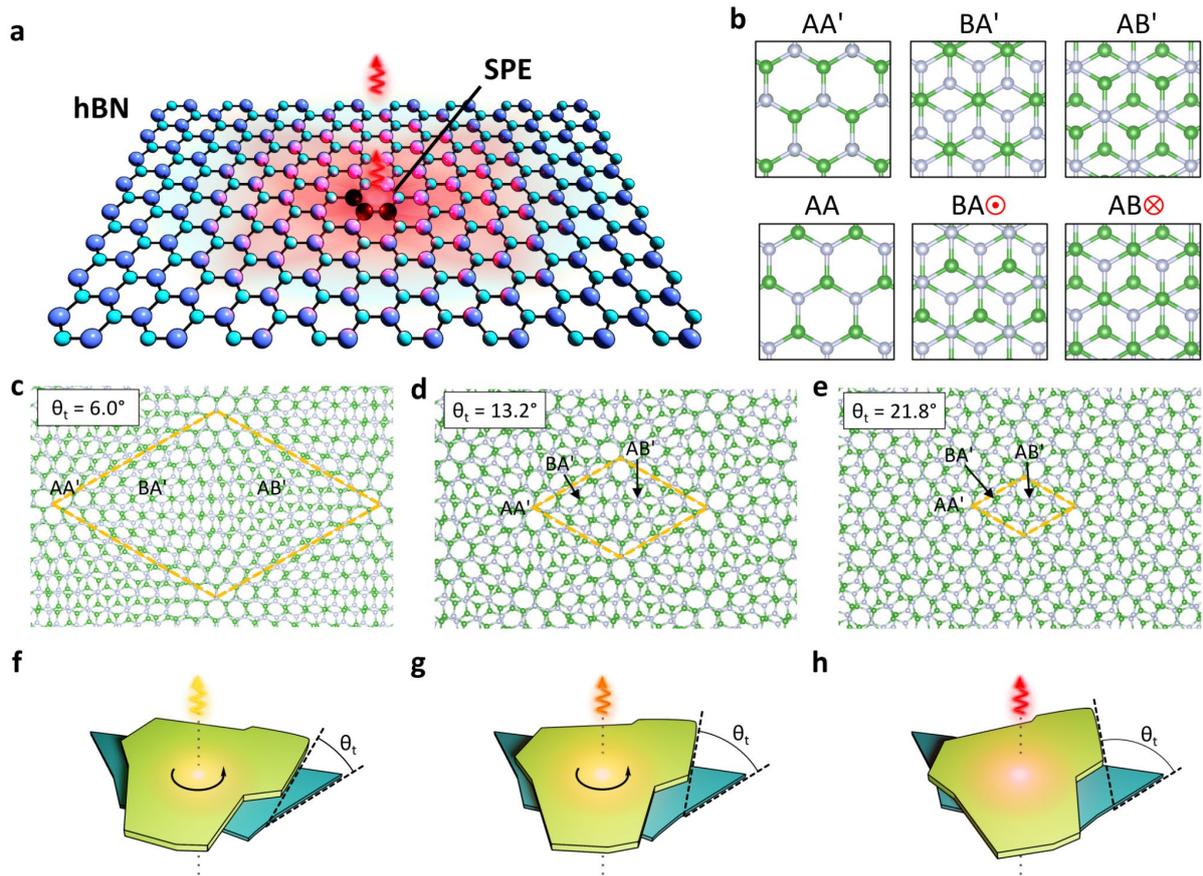

*Figure 1. Modulation of single defect emissions in hBN through twisted interfaces. a) Illustration of an in-plane carbon trimer single photon emitter (SPE) in a single layer of hBN. b) Possible configurations of stacked bilayers of hBN, showing unique stacking orders for an untwisted interface. See main text for details. c-e) The atomic structure of twisted hBN bilayers at (c) $\theta_t$ = 6.0°, (d) $\theta_t$ = 13.2° and (e) $\theta_t$ = 21.8°. The yellow diamond outlines the unit cell of the moiré superlattice. (f-h) Schematic of stacked hBN flakes at different twist angles, $\theta_t$ indicating a shift in the SPE emission energy. The lower flake remains static and top flake is rotated in f-h)*

The degree of atomic reconstruction in twisted hBN also plays a crucial role in the local stacking configuration and is highly dependent on the moiré periodicity. It has been extensively discussed that, for small $\theta_t$, the large moiré periodicity allows for significant reconstruction, expanding energetically favorable domains (e.g., AA' or BA/AB) while shrinking unfavorable regions.[37] Conversely, larger $\theta_t$ values result in smaller periodicities that suppress extensive atomic relaxation, thus offering an alternative mechanism for tuning local stacking.

Figures 1(c-e) show the fully relaxed crystal structures for large-angle superlattices with $\theta_t$ = 6.0, 13.2, and 21.8°, rotating from the AA' configuration, with corresponding moiré periodicities of 23.98, 10.96, and 6.65 Å, respectively. These angles give the commensurate supercell structure which is determined by the hexagonal symmetry, without any additional strain. The lattice exhibits a smooth continuum of atomic configurations that transition between local

environments resembling the ideal AA', BA', and AB' stackings. This establishes the twist angle as an effective physical knob for engineering the local atomic environment.

While the discussion so far has been focused on twisted bilayers, multilayer or bulk hBN flakes can also be stacked and twisted, leading to an interface with a moiré superlattice or corresponding periodic stacking configurations. In this way, it is proposed that controlled tuning of SPEs can be implemented experimentally, based on the relative $\theta_t$ as shown schematically in Figures 1(f-h). There, the emission energy of the SPE at the interface is red shifted across the three angles $\theta_{t1}$ - $\theta_{t3}$.

To investigate how the angle $\theta_t$ affects the emission energy of SPEs in hBN, first-principles DFT calculations were performed. We focus on the carbon trimer defects as a model system that was predicted to be the source of the 2 eV SPEs in hBN.[29, 30] Figures 2(a, d) show the atomic configurations of two distinct carbon trimer defects (indicated by brown atoms) in an hBN lattice, referred to as $C_2C_B$ and $C_2C_N$. Their corresponding electronic structures, calculated at the generalized gradient approximation (GGA) level, are shown in Figures 2 (b,e), respectively. Both defects introduce multiple spinful defect states within the hBN bandgap, with transition energies of 1.192 eV ($C_2C_B$) and 1.204 eV ($C_2C_N$) in AA' stacking configuration, confirming their potential to serve as SPEs. Hereafter, the defect transition energy is denoted as $E(X)^N(\theta_t)$, where X and N represent the defect and the stacking configuration, respectively, at a given $\theta_t$.

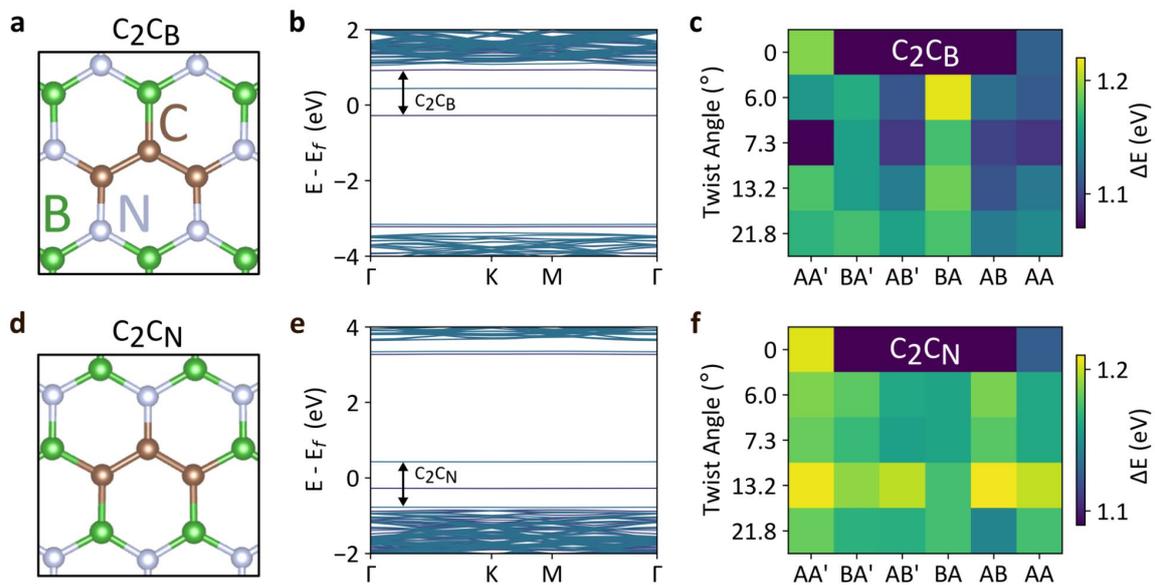

*Figure 2. First principles DFT calculations of carbon trimer defects in hBN. a) Atomic configurations of the $C_2C_B$ carbon trimer defect and b) the corresponding electronic structure in an 8x8 supercell of AA' stacking bilayer hBN. (c) Color coordinate maps illustrating the energy difference between occupied and unoccupied energy levels for the $C_2C_B$ spin-down defect state, with respect to the twist angle and local stacking configuration. d) Atomic configuration of the $C_2C_N$ carbon trimer defect and e) the corresponding electronic structure in an 8x8 supercell of AA' stacking bilayer hBN. (f) Color coordinate maps illustrating the energy difference between occupied and unoccupied energy levels for the $C_2C_N$ spin-up defect state, with respect to the twist angle and local stacking configuration. The blue and purple solid lines in (b,e) represent the spin-up and spin-down states, respectively.*

Figures 2(c, f) illustrate how the calculated transition energies for these defects vary as a function of both the local stacking configuration and $\theta_t$. (Crystal structures and detailed defect transition energies are outlined in Figures S1-4 and Tables S1 and S2 of the supplementary information, respectively). To understand the complex energy variations in the color maps, we begin with our discussion on the untwisted cases. For AA stacking at ($\theta_t = 0$), the transition energies $E(C_2C_B)$ and $E(C_2C_N)$ were calculated to be 1.117 eV and 1.126 eV, respectively, corresponding to redshift of 75 meV and 78 meV compared to their AA' counterparts. This strong stacking dependence is consistent with calculated emission tuning at hBN bilayers, even in the absence of twisting.[38]

Introducing finite $\theta_t$ reveals an even richer landscape for energy variation. For example, $E(C_2C_B)^{AA'}(\theta_t)$ and $E(C_2C_N)^{AA'}(\theta_t)$ are calculated to be 1.187, 1.182, 1.207, and 1.182 eV, and 1.15, 1.071, 1.78, and 1.167 eV, respectively, for $\theta_t$ = 6.01, 7.34, 13.17, and 21.78°, showing a non-monotonic dependence on $\theta_t$. The position of carbon trimer defects within the moiré supercell provides another tuning dimension. At $\theta_t$ = 6.01°, for instance, $E(C_2C_N)^{BA'}(6.01°)$ and $E\_(C_2C_N)^{AB'}(6.01°)$ is redshift by -7 and -25 meV compared to $E(C_2C_N)^{AA'}(6.01°)$. $E(C_2C_B)^{BA'}(6.01°)$ and $E(C_2C_B)^{AB'}(6.01°)$ shift by -13 meV (redshift) and by 37 meV (blueshift), respectively.

Additional insights comes from comparing BA and AB stacking configurations, which are mirror images of each other at $\theta_t = 0$. Hence, their energy difference directly captures the effect of the out-of-plane moiré dipole. We found that $E(C_2C_N)^{BA}(\theta_t)$=1.16, 1.16, and 1.174 eV for $\theta_t$ = 6.01, 7.34, and 13.17°, consistently redshifted by -34, -19, and -26 meV compared to $E\_(C_2C_N)^{AB}$ at the same $\theta_t$. In contrast, $E(C_2C_B)^{BA}(\theta_t)$=1.214, 1.176, and 1.187 eV at the same $\theta_t$, showing a consistent blueshift by 78, 76, and 90 meV from $E(C_2C_B)^{AB}$. These results highlight that the out-of-plane moiré dipole tunes the wavelength of SPEs by up to 100 meV.

Our findings imply several critical conclusions for engineering SPEs in twisted hBN. First, emission energies are controlled not only by $\theta_t$ but also by the emitter's spatial position within the moiré supercell. Second, while the out-of-plane moiré dipole provides a clear tuning effect, the strong energy modulation in non-polar regions (AA, BA', AB') suggests that the local atomic environment is also a determining tuning factor. While *a priori* prediction of the exact emission energy is challenging, the twist angle can be used to generate a broad spectrum of quantum emission energies from identical defects. This provides a new and powerful degree of freedom for manipulating light at the nanoscale and designing hBN-based SPEs.

To validate the proposition of the twist-controlled tuning of emission energy of SPEs as shown above, we engineered a homobilayer of hBN crystals that were mechanically twisted. In our scheme, one flake (bottom one) hosts the SPEs and another one (top) creates a moiré superlattice at the interface that can be dynamically modulated. In essence, the top hBN flake acts as a lever to modulate the embedded SPEs in the bottom flake.

To generate the SPEs, hBN flakes were exfoliated and annealed in an oxygen atmosphere at 1000 °C. To ensure proximity of SPEs to the twisted interface, thin hBN flakes < 20 nm were first identified based on optical contrast and colour using optical microscopy. A typical thin flake with a thickness of ~15 nm is shown in Figure 3a, and a corresponding atomic force microscope (AFM) image and line scan are shown in Figure 3b.

Photoluminescence (PL) measurements carried out at room temperature under ambient conditions confirmed the presence of multiple emitters. Confocal PL maps of this hBN flake are shown in Figures 3 (c,d), outlining the position of two isolated SPEs, labelled E1 and E2, respectively. A confocal PL map of the entire 25 µm x 25 µm region can be seen in Figure S6. The spectra of these and other typical emitters from the hBN flake are shown in Figure 3e. Multiple narrowband emitters with zero phonon lines (ZPLs) ranging from 590 - 760 nm (1.6 - 2.1 eV) are observed. Finally, to ascertain the quantum nature of the emitters, second order correlation measurements from E1 and E2 are shown in Figures 3(f, g). The dips at zero delay time with $g^2(0) < 0.5$ indicating they are SPEs.

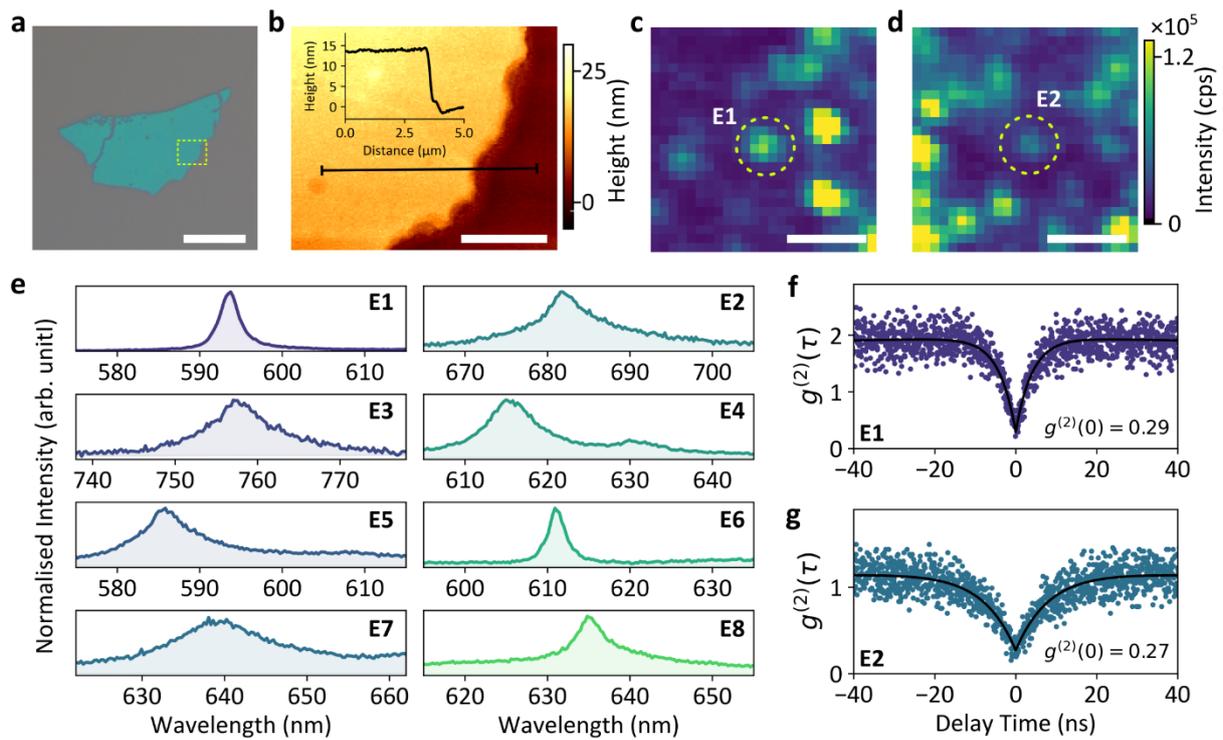

*Figure 3. Engineering quantum emitters in thin hBN. a) Optical image of an hBN flake on SiO$_2$/Si substrate. The dotted box corresponds to the imaged AFM region in b). The scale bar corresponds to 20 µm. b) AFM linescan of the hBN flake in a). Inset is a line profile taken along the black line. The scale bar corresponds to 2 µm. c,d) Confocal PL maps from the hBN flake, showing localised and isolated spots corresponding to SPEs. The circled regions show the locations of two representative SPEs, E1 and E2. The scale bars correspond to 2 µm. e) Individual spectra collected from multiple sites E1-E8 hosted in the hBN flake shown in a). f,g) Second order correlation function plots from SPEs E1 and E2, respectively.*

After identifying a thin hBN flake hosting SPEs, a second flake of pristine hBN, was transferred via a dry polymer stamp based method. In our case, dynamic tuning of the top hBN is important to directly check the effect of multiple twist angles, $\theta_t$ on the SPEs. Note, that this step is crucial and is fundamentally different from the common practice in the field of vdW twistronics where multiple twist angles are achieved by simply making different samples.[27] Indeed, for hBN in particular, the twist angle between the top and bottom crystals is often locked after transferring due to the strong adhesion via VdW forces. Notably, using the twisting procedure developed below, the top hBN can be reconfigured, allowing for multiple $\theta_t$ from the same sample.

The workflow is schematically outlined in Figure 4a. We employ a hemispherical polydimethylsiloxane (PDMS) stamp on a glass with a polyvinyl alcohol (PVA) sticking layer. To ensure sufficient adhesion to the stamp during the pick-up step, a temperature of 55 °C was maintained using a heating stage. Once the top hBN flake is detached and adheres to the stamp, the bottom flake can then be rotated using a rotation stage. The angular precision is on the order of 1°, however, finer control could be implemented using a micrometer equipped stage. When the rotation is complete, the top flake is lowered and brought back in contact with bottom hBN. The temperature is increased to 120 °C to melt the PVA and release the top flake. Residual PVA polymer is removed using water immersion. Complete details of the twisting procedure are discussed in the methods section. After each rotation, the optical measurements are repeated, as will be shown below.

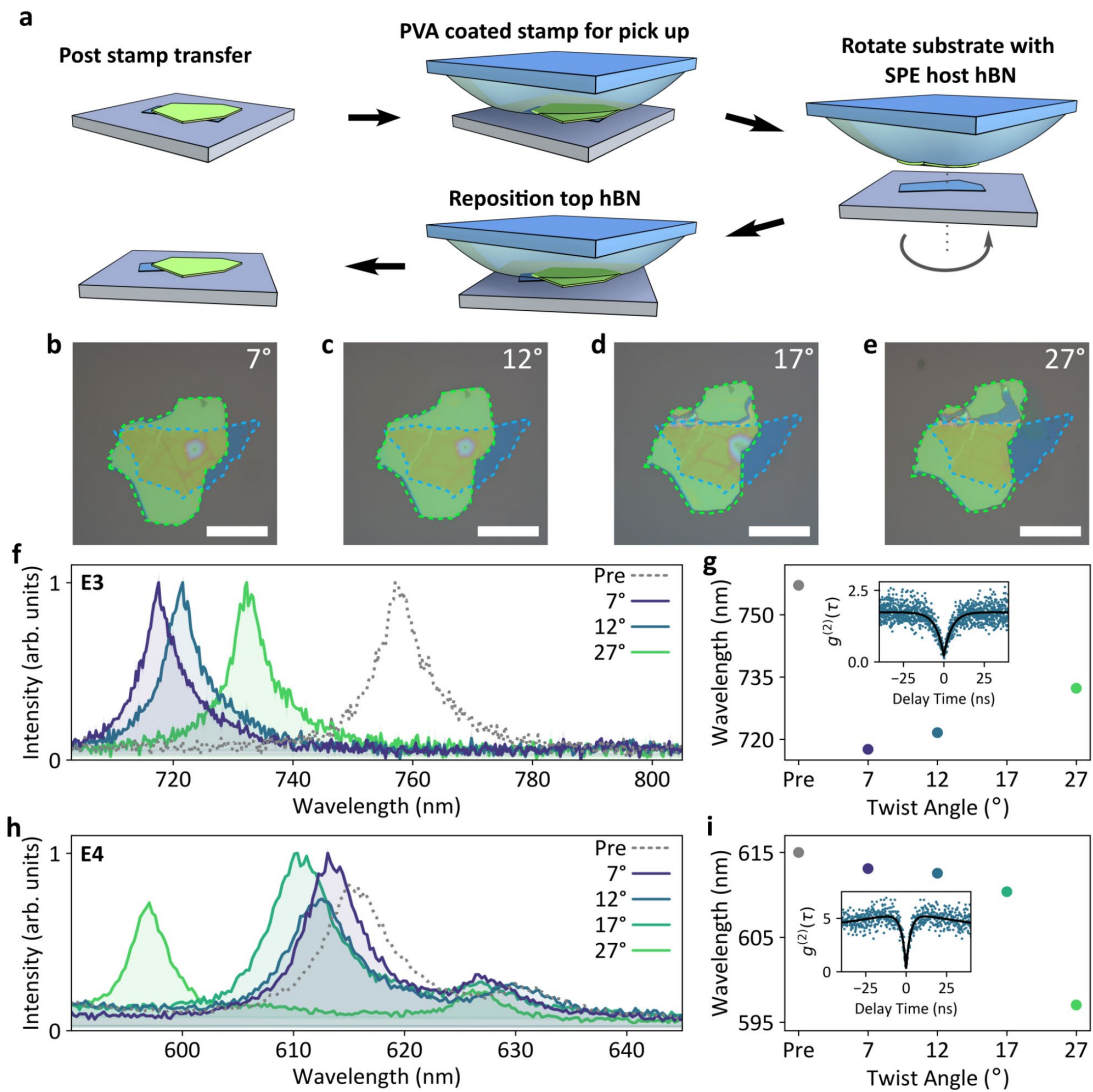

*Figure 4. Experimental observation of twist-controlled tuning of quantum emitters in hBN. a) Schematic of the stamp-based twist procedure. Beginning with an initial stack of hBN (Post stamp transfer), a PVA coated PDMS stamp is used to pick up the top hBN flake. While the top hBN remains on the stamp, the substrate with lower hBN can be*

*rotated via a rotation stage. The top hBN is then placed back down on the hBN at a new $\theta_t$. This procedure can be repeated as needed. b-e) Optical images of the stacked hBN flakes at $\theta_t$ = 7°, 12°, 17° and 27°. f) PL spectra of E3 at each twist angle. Note that no emission was visible at $\theta_t$ = 17°. g) Peak values of E3 as a function of $\theta_t$ (twist angle). (inset) Second order correlation at $\theta_t$ = 27°. h) PL spectra of E4 at each twist angle. i) Peak values of E4 as a function of twist angle. (inset) Second order correlation at 27°. "Pre" labels in f-i) refer to measurements taken on the bottom flake only, before the top hBN flake was added.*

Our developed method is facilitated by the annealing process of the bottom hBN flake that increases the adhesion to the $SiO_2$/Si substrate and ensures that only the top flake is picked up during the twisting process. The interface also remains clean as there is only hBN to hBN contact, and there is no requirement for any specific size, orientation or flake geometry which may be required in other angle controlled methods.[39] The use of the PVA polymer also ensures that background fluorescence from any residual polymer is minimised (See Fig S5). Figures 4(b-e) show the same hBN mono-bilayers in four $\theta_t$ configurations ($\theta_t$ = 7, 12, 17 and 27), by repeating the twisting procedure above. To measure the absolute $\theta_t$ between the hBN flakes, electron backscattered diffraction (EBSD) measurements were undertaken (Fig S5). Additional samples were prepared using similar methods, as shown in the supporting information (Fig S7).

We now turn to characterise the same SPEs after every twist, namely at $\theta_t$ = 7, 12, 17 and 27°. Figures 4 (f, h) outline the spectral shift of two SPEs, E3 and E4 at each $\theta_t$, and show their corresponding second order autocorrelation functions confirming that quantum behaviour is maintained all throughout. For emitter E3 (Figure 4f,g), the ZPL emission is first blue-shifted by > 30 nm (~100 meV), at $\theta_t$ = 7° when compared to the unstacked hBN flake. However, at each successive twist $\theta_t$ the ZPL emission then red shifts, as seen in Figure 4f. Figure 4g displays the peak ZPL as a function of the twist angle, together with the second order correlation measurement of the same emitter after the final twist of $\theta_t$ = 27°.

A slightly different behaviour is observed for emitter E4. Here, a monotonic blue shift is recorded for this emitter, as $\theta_t$ is gradually increasing. A total shift of ~ 20 nm (~ 80 meV) is observed, as seen in Figure 4(h, i). Similarly to E3, the quantum behaviour of this emitter is maintained as evident by the $g^{(2)}(\tau)$ in Figure 4i. Additional spectra from other sites E1-E8 are presented in Figure S9.

At first glance these results appear mixed, however the observed variation of emissions in both magnitude and direction are not unexpected. A linear or otherwise ordered trend in the emission energy shift is not predicted for any single lattice stacking across all the simulated angles, $\theta_t$ (Fig. 2(c,f)) - namely for the entire span of twist angles a continuous blue (red) shift for any stacking is not expected. However, the calculated and experimental $\theta_t$ outlined in this work are relatively large, resulting in small moiré superlattice unit cells of on the order of few to tens of nanometers (see Figure 1). This means that it is highly probable that the localised stacking orders are not maintained for each $\theta_t$, but rather alternate from configuration to configuration. Given the rich variety of configurations (six as shown in Figure 1b), this is highly likely and explains well the shifts observed for E3 and E4. For example, the large emission blue-shift for E3, compared to the unstacked spectrum (Labelled "Pre" in Figure 4(f)) to the first stacking angle $\theta_t$ = 7°, is indicative of a stacking modulation to a higher energy BA order for the $C_2C_B$ defect (Calculated $\theta_t$ = 6° from Fig. 2(c)). The gradual red-shift with increasing $\theta_t$ is also expected when the defect remains in the same BA stacking order for the calculated $\theta_t$ = 13.2 and 21.8°.

The magnitude of the emission shifts is also expected to vary, but would likely depend on how far the emitter is located from the interface. The modelling in Figure 2 outlines the emission shifts for a defect at the interface. However, a shift in emission intensity is still possible for an SPE embedded a few nm below the interface, which is why we engineered the emitters in a thin hBN flake of ~ 15 nm.

Finally, there are important nuances to address with regards to the modulation mechanism. Firstly, the twisted interfaces induce an out of plane electric polarization at AB/BA stacking sites, independent of any defect based SPEs in the lattice. The electrostatic potential, $V$ is dependent upon both the vertical position, $z$ and lateral position, R as well as the moiré period, $b$. At an AB/BA stacking site, the potential magnitude $|V|$ can be approximated as follows,[40]

$$|V(z)| \approx sgn(z)\frac{P}{2\epsilon_0}e^{-G|z|} \qquad (1)$$

With $G = 4\pi/\sqrt{3}b$, electrical polarisation, $P$ = 2.25 pC/m taken from Reference [36] and vacuum permittivity, $\epsilon_0$. $|V|$ is largest when both $z$ is minimised, and $b$ is maximised. Such conditions are typically found at small twist angles of $\theta_t$ < 5°.[3, 40] When considering a point-like SPE in a close proximity to the twisted interface ($z \cong$ 0.5 nm), we do expect a non-zero shift of the energy levels, due to the significant $|V|$, even at large $\theta_t$ > 5° (See Figure S10). However, given the in-plane dipole orientation of the measured SPEs as confirmed by polarisation measurements (See Figure S11), the maximum shift of the emission should only be on the order of a few meV at best[41, 42], orders of magnitude lower than the large experimental shifts noted for E3 and E4.

Secondly, tuning of SPE emission in hBN has before been demonstrated via strain effects[43]. As the SPEs in our twisted system are hosted solely within the bottom flake of hBN, and are not subject to strain effects, this can be ruled out as a possible influence. Further, any residual strain due to pressure from the stamp would be applied vertically and unlikely to affect emitters that are polarised in plane. We do note some folding along grain boundaries, however, all optical characterisation is undertaken after this step on emitters far from the boundaries.

Finally, recent studies have demonstrated an enormous increase in emission at the UV spectral range from a twisted hBN.[21, 27, 28] While originally ascribed to defect emissions, it became apparent that the enhanced brightness was associated with a formation of self trapped excitons when excited via high-energy electrons.[21] These emissions are broadband and confined to the UV region, without any localised or quantum emitters. Our current study focuses on the tuning of narrowband SPEs in the visible range that are excited with a deep wavelength laser (λ=532 nm) that does not require or result in a formation of any excitons.

To conclude, we have demonstrated that the twist degree of freedom in a van der Waals homobilayer of hBN provides a unique platform for control emission properties of isolated quantum systems at room temperature. To this end, we developed a stamp-based method that overcomes strong interlayer adhesion and enables fine twist-angle control, allowing in-situ mechanical rotation of the hBN homobilayer. Utilising this approach, we mechanically modulated SPEs across multiple twist angles, $\theta_t$ to show both blue and red shifts up to ~30 nm (~100 meV), while maintaining single photon purity. First principles calculations for carbon trimer defects ($C_2C_B$ and $C_2C_N$) are consistent with these observations, indicating that the relevant defect transition energies are strongly sensitive to both $\theta_t$ and the local stacking order within the moiré lattice.

Our work opens up exciting avenues towards programmable quantum photonics architectures with van der Waals materials. For instance multiple emitters with distinct emission ZPLs could be potentially brought into resonance using a mechanical twist, thus leading to a realisation of spatially programmable emitter arrays defined by the moiré lattice. Further, integrating nano electromechanical actuators[23] together with already established hBN nanophotonic elements[44], such as waveguides and cavities, will enable a tunable and programmable vdW quantum photonic circuitry. Finally, correlating emission with local stacking orders using a combination of advanced Transmission Electron Microscopy techniques combined with scanning tunneling microscopy and cathodoluminescence imaging can further shine light on the nature of these localised quantum emitters.[45, 46] All in all, the in-situ twist control provides an elegant and compact material-intrinsic degree of freedom for modulating solid state quantum systems at the atomic scale at vdW systems.

## Methods
### First-Principles Calculations

First-principles calculations based on density functional theory (DFT) were performed using the Vienna *ab initio* simulation package (VASP). We employed the projector-augmented wave (PAW) pseudopotentials and the generalized gradient approximation (GGA) of the Perdew-Burke-Ernzerhof (PBE) exchange-correlation functional. The kinetic energy cutoff for the plane wave basis was set to 500 eV. All crystal structures were fully relaxed with a force criterion of 0.01 eV/Å, and van der Waals (vdW) interactions were treated using the Grimme-D3 method.

The in-plane lattice constant of hexagonal boron nitride (hBN) was calculated to be 2.514 Å. We then constructed supercell structures for AA and AA' stacked bilayer hBN with twist angles of 0°, 21.78°, 13.17°, 7.34°, and 6.01°. The corresponding supercell lattice constants were determined to be 2.514, 6.650, 10.956, 19.631, and 23.977 Å, respectively. To describe carbon trimer defects, we used 8x8, 3x3, and 2x2 supercell structures for the 0°, 21.78°, and 13.17° twist angles, respectively. Then, $C_2C_B$ and $C_2C_N$ were placed at various sites within the supercells, to examine the influence of the local moiré environment on emission properties. All defect configurations considered in this study are presented in Figures S1-4, and the corresponding Brillouin zones were sampled by Γ point.

### Sample Preparation

hBN crystals were grown via an atmospheric pressure, high temperature method. For full details see Figure S12 in the supplementary information. Pristine and carbon-doped hBN flakes were exfoliated via the scotch tape method onto separate 285 nm $SiO_2$/Si substrates. After exfoliating, carbon-doped hBN was annealed for 4 hours at 1000 °C in an oxygen rich atmosphere (1000 sccm $O_2$ flow), followed by an ozone clean in a commercial UV ozone system. Pristine hBN was used as the top stacking layer and was not annealed before being transferred. hBN flake thickness was measured with a Park Systems XE7 AFM using a 160AC-NA tip in non-contact mode.

### Stamp-Based Twisting Procedure

A hemispherical PDMS stamp was prepared by drop-casting the two-part solution onto a glass slide at 180 °C to form a dome. Once solidified, PVA solution was drop-cast onto the PDMS surface at 50 °C and allowed to dry. The PDMS/PVA stamp was mounted on a holder with XYZ three-axis stage, and the target hBN on $SiO_2$/Si substrate was placed on a temperature-controlled rotation stage with Z micro actuator (Thorlabs). For the pick-up, the bottom of the stamp was slowly brought into contact next to the target hBN flake. The stage temperature was set to 55 °C to increase the contact area, and once fully covered by the stamp, was retracted to lift off the flake. For the initial twist angle, the substrate was removed once the top hBN was attached to the stamp and then replaced by the substrate with SPE host hBN. For all additional angles the same substrate was left on the rotation stage. The stage was subsequently rotated to the desired twist angle ($\theta_t$), and lateral alignment between the flakes was refined using the three-axis stage. For release, the stamp with top hBN flake was slowly lowered until contact was achieved between both hBN flakes. After contact, the stage temperature was increased to 120 °C to melt the PVA and promote transfer. The PDMS lens was retracted, leaving the PVA film on the substrate together with the transferred top flake. Residual PVA was removed by immersing the sample in deionized water at 60 °C, followed by rinsing and $N_2$ drying. This pick-up/rotate/release sequence was repeated as needed to realize multiple $\theta_t$ on the same device; all temperatures refer to the sample stage and all operations were performed under ambient conditions.

### Optical Measurements

Photoluminescence measurements were performed on a lab-built confocal microscope with 532 nm continuous-wave (CW) excitation laser and 100x objective (Nikon 0.9 NA). A 532 nm long-pass dichroic mirror and 568 long-pass filter were used to filter any reflected laser emission. Confocal mapping was undertaken with a 4f system with a fast steering mirror (Newport FSM-300). Spectra were collected with a fiber coupled spectrometer (Princeton Instruments) and 300 lines/mm grating. Second order correlation measurements used a 5050 beamsplitting fiber and two APDs (Excelitas SPCM). Additional filters were used to isolate SPEs and reduce background emissions for Figure 4g,i.

**Acknowledgements** We acknowledge financial support from the Australian Research Council (CE200100010, FT220100053, DP240103127, DP250100973), the Air Force Office of Scientific Research under award number FA2386-25-1-4044 and the UTS node of the ANFF for access to nanofabrication facilities. This research was supported by the National Research Foundation of Korea (NRF) grant funded by the Korean government (MSIT) (2021R1A2C2093155, 021M3H4A1A03054856, 2022M3H4A1A04096396, RS-2023-00254055). This work was partly supported by the Institute of Information & Communications Technology Planning & Evaluation(IITP)-ITRC(Information Technology Research Center) grant funded by the Korea government(MSIT) (IITP-2025-RS-2024-00437191).